\documentclass[aip, jap, amsmath, amssymb, reprint]{revtex4-2}

\usepackage{hyphenat}
\usepackage[pdftex]{graphicx}% Include figure files
\usepackage[american]{babel}
\usepackage{booktabs} % Horizontal rules in tables
\usepackage{float} %fix image position with [H]
\usepackage{dcolumn}% Align table columns on decimal point
\usepackage{bm}% bold math
\usepackage[svgnames]{xcolor}
\usepackage{caption}
\usepackage{subcaption}
\usepackage{braket}
\usepackage{notes2bib}
\usepackage{tikz}
\usepackage{comment}
\usepackage{nicefrac}
\usepackage[normalem]{ulem}

\newcommand\CORRremove[1]{}

\newcommand\CORRBremove[1]{}

\usepackage[svgnames]{xcolor}

\usepackage[normalem]{ulem}
\usepackage{lipsum}

\begin{document}
	\selectlanguage{american}

	\title{Carrier Thermalization and Biexciton Formation in a Polar ZnO/Zn$_{\mathbf{0.84}}$Mg$_{\mathbf{0.16}}$O Quantum Well Probed by Ultrafast Broadband Spectroscopy}
	
	\author{Daniel O. Siebadji Tchuimeni}
	\email{daniel-osee.siebadji-tchuimeni@etu.unistra.fr}
	\affiliation{IPCMS UMR7504 CNRS - université de Strasbourg
		23, rue du L{\oe}ss B.P. 43 - F-67034 Strasbourg CEDEX 2 - France
	}%
	\affiliation{
		Université de Toulouse , INSA-CNRS-UPS,LPCNO ,135 avenue de Rangueil, 31077 Toulouse France
	}%
	
	\author{Marc Ziegler}
	\author{Olivier Crégut}
	\author{Pierre Gilliot}
	\affiliation{IPCMS UMR7504 CNRS - université de Strasbourg
		23, rue du L{\oe}ss B.P. 43 - F-67034 Strasbourg CEDEX 2 - France
	}%
	\author{Christian Morhain}
	\affiliation{
		Université Côte d’Azur, CNRS, CRHEA	Rue B. Grégory, Valbonne 06560, France
	}%
	\author{Andrea Balocchi}
	\affiliation{
		Université de Toulouse , INSA-CNRS-UPS,LPCNO ,135 avenue de Rangueil, 31077 Toulouse France
	}%
	
	\author{Mathieu Gallart}
	\email{mathieu.gallart@ipcms.unistra.fr}
	\affiliation{IPCMS UMR7504 CNRS - université de Strasbourg
		23, rue du L{\oe}ss B.P. 43 - F-67034 Strasbourg CEDEX 2 - France
	}%

	\date{\today}% It is always \today, today,
	\begin{abstract}{\tiny }
		We investigate the ultrafast dynamics of excitons in a 2.6 nm-thick $\mathrm{ZnO/Zn_{0.84}Mg_{0.16}O}$ quantum well grown on a \textbf{c}-axis sapphire substrate, using non-degenerate time-resolved pump–probe spectroscopy. A pump pulse at 266 nm generates photocarriers within the ZnMgO barriers, and their dynamics is monitored through time-resolved differential reflectance measurements using a supercontinuum probe spanning the 345–400 nm spectral range. Photocarriers generated in the barriers rapidly relax into the quantum well, where they form excitons within sub-picosecond timescales. These excitons quickly thermalize and become localized, likely due to interface disorder or well-width fluctuations, as supported by photoluminescence measurements showing a clear Stokes shift and the absence of free exciton emission. A phonon-assisted absorption process, leading to the effective thermalization of excitons, is observed and analyzed. We identify moreover a negative differential reflectance feature as a photoinduced absorption into a biexciton state, with a binding energy ranging from 18 to 22 meV depending on temperature.
	\end{abstract}

	\maketitle

	%%%%%%%%%%%%%%%%%%%%%%%%%%%%%%%%%%%%%%%%%%%%%%%%%%%%%%%%%%%%%%
	\section{Introduction}
	Semiconductors with a wide bandgap have long attracted significant interest due to their applications in optoelectronic devices operating in the short-wavelength regime. Among them, ZnO offers the potential for functional applications even at room temperature, owing to its large exciton binding energy of 60 meV~\cite{Capper2011, Klingshirn2007}, along with other attractive electronic properties~\cite{Klingshirn2007}. Some of the key applications include solar cells~\cite{Vittal2017}, light-emitting diodes~\cite{Ryu2006}, optically and electrically pumped ultraviolet lasers~\cite{Chu2008, Hofstetter2007}, and spintronic devices~\cite{Pearton2006}, among others. These applications are primarily explored under controlled conditions within heterostructures such as quantum wells (QWs), where quantum confinement effects become well-defined as a function of well width~\cite{Morhain2005}.
	
	Among these systems, wurtzite ZnO epitaxial quantum wells grown along the \textbf{c}-axis exhibit an internal electric field due to the combined effects of piezoelectric and spontaneous polarization fields, which are determined by the barrier material composition~\cite{Bretagnon2007}. Through the quantum-confined Stark effect (QCSE), this internal electric field strongly influences the optical properties of ZnO QWs by significantly modifying the excitonic transition energies and oscillator strengths. Furthermore, the symmetry breaking along the \textbf{c}-axis induced by the QCSE introduces a Bychkov-Rashba term in the spin-orbit Hamiltonian. In appropriately engineered structures, this effect can be exploited to compensate for the Dresselhaus term, thereby suppressing the dominant Dyakonov-Perel spin relaxation mechanism for electrons~\cite{Harmon2011}.
	
	Excitonic complexes are bound states formed by multiple excitons or combinations of excitons with other charged particles within a semiconductor. These complexes arise due to Coulombic interactions and play a crucial role in the optical and electronic properties of materials, particularly in nanostructures, where their binding energies are significantly enhanced compared to their bulk counterparts~\cite{Charbonneau1988,Yamada1995}. Since their first observation in CuCl~\cite{Mysyrowicz1968}, biexcitons have been identified in several bulk materials, including GaAs~\cite{Brinkman1973}, GaN~\cite{Kawakami1996}, and ZnO~\cite{Ko2000,Yamamoto2001,Hazu2003}, as well as in quantum wells~\cite{Charbonneau1988,Yamada1995}. In ZnO quantum wells, biexciton recombination has been observed in the photoluminescence (PL) spectra of both polar~\cite{Sun2001,Sun2002} and non-polar~\cite{Ali2018} QWs.
	
	In this context, understanding the formation and relaxation dynamics of excitonic complexes is essential, not only from a fundamental perspective but also for evaluating the potential of these systems in spintronic applications. Indeed, the strong Coulomb interactions that characterize ZnO-based heterostructures can lead both to the rapid formation of bound states—such as excitons and biexcitons—and to their localization, either due to well-width fluctuations or interface disorder. While these effects contribute to the robustness of excitonic features even at elevated temperatures, they may also enhance exchange interactions between carriers and accelerate spin depolarization processes.
	
	In the present work, we use ultrafast pump–probe spectroscopy to investigate how strong Coulomb interactions drive the thermalization, localization, and complexation of photogenerated carriers in a ZnO/ZnMgO quantum well, and how these processes may ultimately influence spin lifetimes. In particular, we explore the interplay between exciton formation dynamics and many-body effects, such as biexciton creation and phonon-assisted transitions, as a means to assess their impact on the temporal and spectral characteristics relevant for spintronic functionalities.
	
	\section{Samples and experimental setups}
	The sample studied is a $\mathrm{ZnO/Zn_{0.84}Mg_{0.16}O}$ double quantum well (QW) structure grown on a \textbf{c}-axis sapphire substrate by molecular beam epitaxy (MBE). A 1~\textmu m-thick ZnO buffer layer was directly deposited on the sapphire substrate, followed by the successive growth of a 200~nm-thick $\mathrm{Zn_{0.84}Mg_{0.16}O}$ barrier, a 2.6~nm-thick QW, another 200~nm-thick $\mathrm{Zn_{0.84}Mg_{0.16}O}$ barrier, a second 7.1~nm-thick QW, and finally a 200~nm-thick $\mathrm{Zn_{0.84}Mg_{0.16}O}$ capping layer.
	
	For the non-degenerate pump-probe experiments, the laser source is a home-built Titanium:Sapphire oscillator generating 800~nm pulses with 80~fs duration and 80~MHz repetition rate. A regenerative amplifier (Coherent REGA 9000) operating at 200~kHz increases the pulse energy to 5~$\mu$J. The third harmonic of the amplified pulses serves as the pump beam at 266~nm (4.66~eV), while the second harmonic is used to generate a supercontinuum probe. This is achieved by tightly focusing the residual 400~nm light onto a 5~mm-thick sapphire crystal using a lens, inducing strong spectral broadening via self-phase modulation, resulting in a supercontinuum spanning a broad spectral range~\cite{Alfano1970}. For our measurements, we select the 340–400~nm range.
	
	Both the pump and probe beams are modulated at different frequencies, $f$ and $2f$ respectively, with $f = 80$ Hz, using mechanical choppers. This modulation scheme ensures that for each on/off step of the pump, the probe has a corresponding exposure and acquisition time, effectively eliminating background noise in the spectrometer detection. The two choppers are synchronized using a modulator. The energy per pulse is 2.7~nJ for the pump and 0.1~nJ for the probe. The estimated pulse duration is slightly above 300~fs, which sets the temporal resolution of our experiment at approximately 150~fs, as it is typically given by half the pulse duration. Based on the corresponding beam spot sizes, we estimate the incident photon fluxes to be in the range of $4.6 \times 10^{12}$ to $4.6 \times 10^{13}$~cm\(^{-2}\) for the pump and $2.4 \times 10^{12}$~cm\(^{-2}\) for the probe.
	
\begin{figure}[h!]
		\includegraphics[width=8cm]{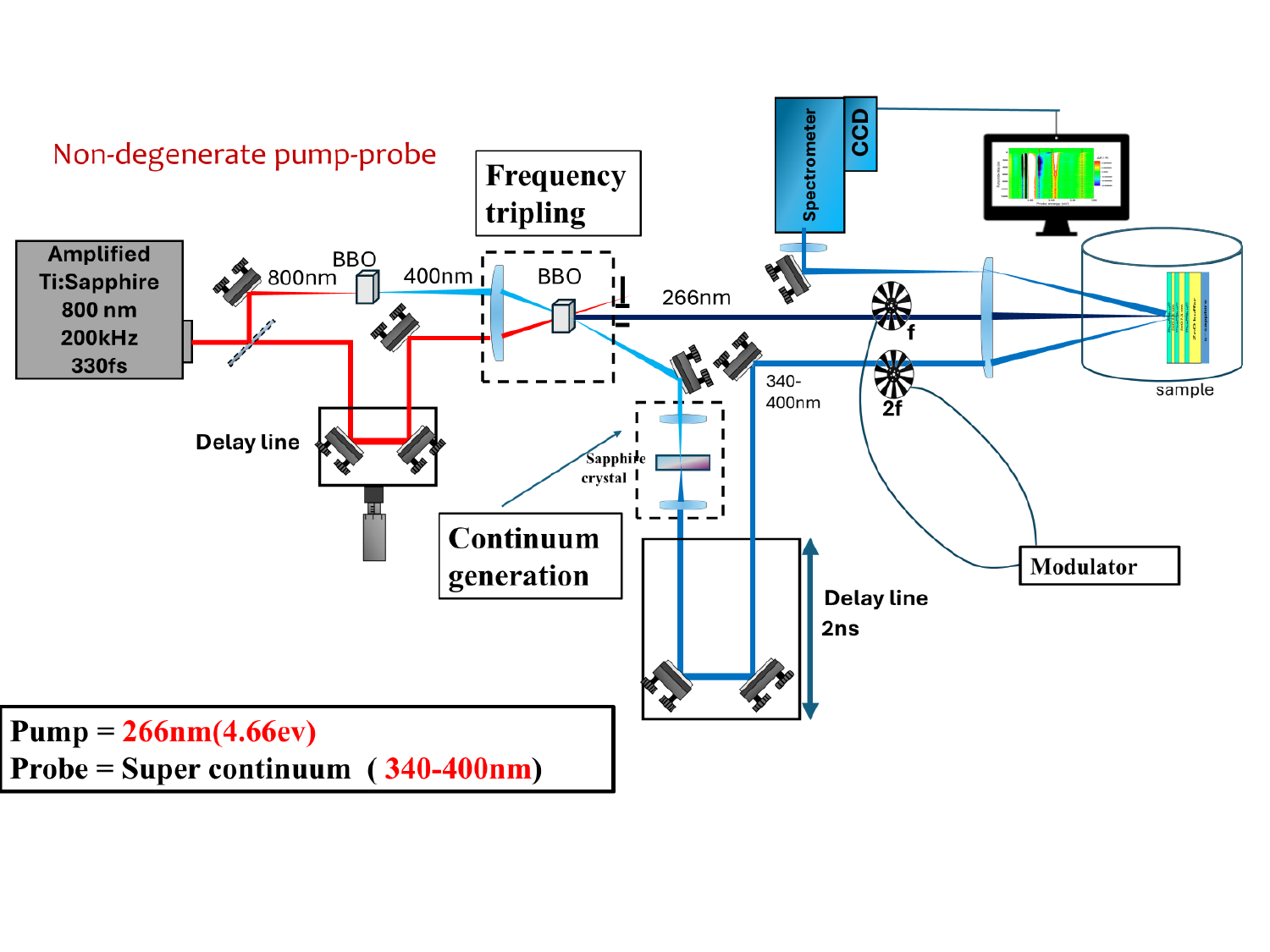}
		\caption{\label{fig:setup} Schematic of the experimental set-up showing pump and probe generation.}
	\end{figure}

	All experiments were performed at low temperature in an Optidry cryostat (\copyright MyCryoFirm). The probe signal is detected using a spectrometer coupled to a cooled charge coupled device (CCD) camera. As both the pump and probe beams are modulated, this setup allows for the direct computation of the differential reflectance spectrum (\(\Delta R/R\)), as illustrated in Figure~\ref{fig:setup}.

	\section{Experimental Results and discussion}
	\subsection{Sample caracterization}\label{Sample caracterization}
	\begin{figure}[h!]
		\includegraphics[width=8cm]{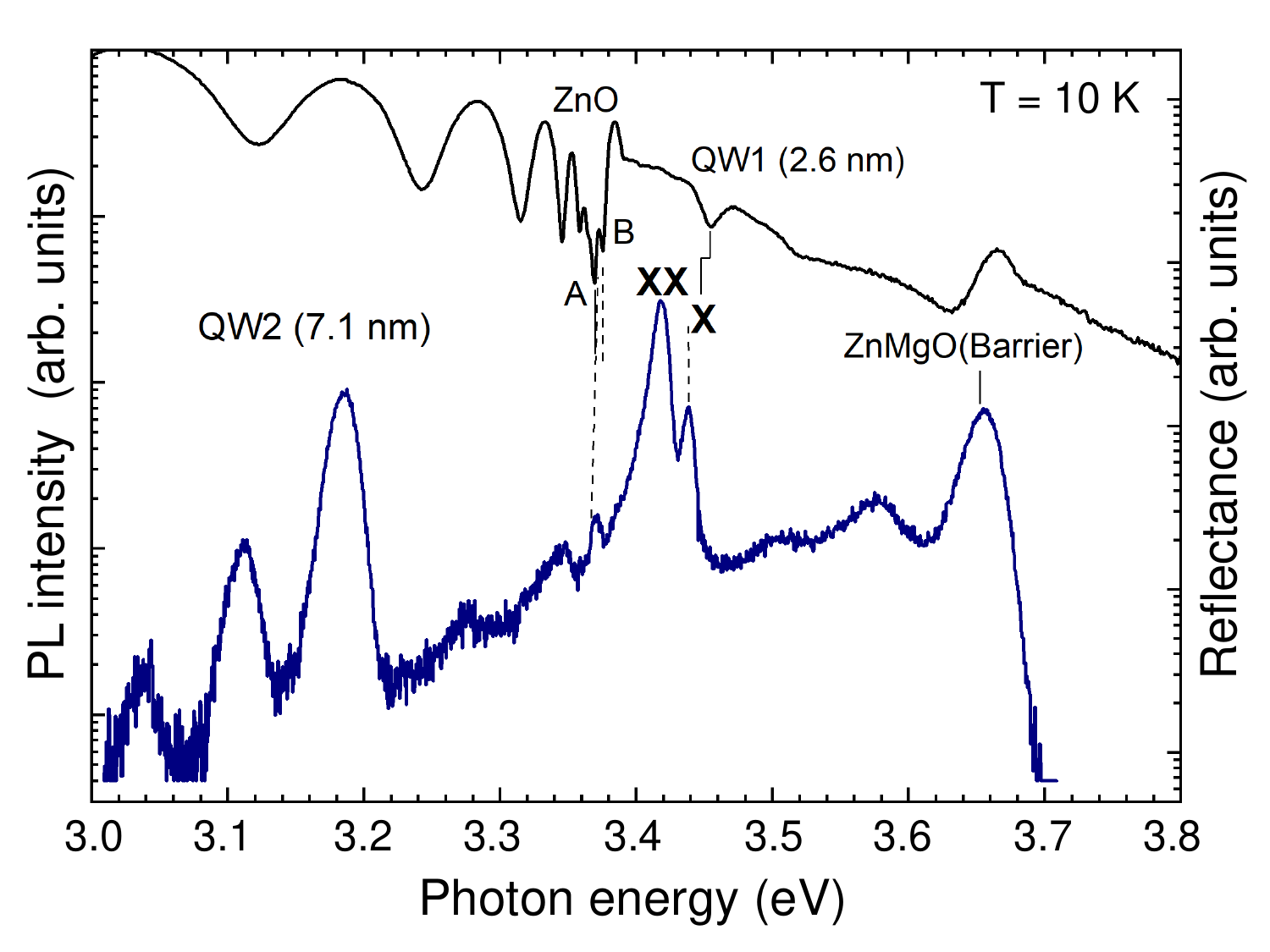}
		\caption{\label{fig:PL} Reflectance (Black line) and PL (Blue line) spectra  of the ZnO/ZnMgO double QW at T = 10 K.}
	\end{figure}
	Figure~\ref{fig:PL} presents the reflectance and photoluminescence spectra of the sample. In the reflectance spectrum, we observe signatures corresponding to the barrier (3.66~eV), the ZnO A and B excitons in the buffer layer (3.36~eV), as well as the heavy-hole exciton (e$_1$hh$_1$) confined in the narrowest well at 3.455~eV.
	
	The wide quantum well does not exhibit any transition in the reflectance spectrum due to the drastic reduction of its oscillator strength by the QCSE. Additionally, a much weaker structure is observed at 3.52~eV, the nature of which will be discussed in the section \ref{Exciton lifetime} as it gives rise to a much more pronounced signature in the $\Delta R/R$ spectrum.
	
	The photoluminescence (PL) spectrum, in turn, reveals recombination lines associated with the ZnMgO barrier and the 7.1~nm QW, both accompanied by their respective phonon replicas. The energy of these replicas is consistent with the optical phonon energy in ZnO ($E_{\mathrm{LO}} = 72$~meV). In the following discussion, we will no longer consider the 7.1~nm QW, as it does not contribute to the optical response in our pump-probe experiment.  
	
	The emission from the 2.6~nm QW exhibits a Stokes shift of approximately 15~meV relative to its reflectance signature, indicating that the PL originates from the recombination of localized excitons in regions where the QW thickness probably locally exceeds its nominal value by one monolayer (ML).
	
	To support this interpretation, calculations based on the envelope function formalism were performed. It is well known that in such structures, the electric field $ F_{\scriptscriptstyle W}$ inside the QW depends on both the well and barrier widths, $L_{\scriptscriptstyle W}$ and $L_{\scriptscriptstyle B}$, respectively, according to the expression~\cite{Leroux1999,Bernardini1998}:
	
	\begin{equation}
		F_{\scriptscriptstyle W}=\frac{L_{\scriptscriptstyle B}}{L_{\scriptscriptstyle W}+L_{\scriptscriptstyle B}}\,F_{\mathrm{max}},
	\end{equation}
	where the dependence of $F_{\mathrm{max}}$ on the magnesium concentration $ x_{\mathrm{Mg}}$ in the barrier is given by~\cite{Bretagnon2007}:
	\begin{equation}
		F_{\mathrm{max}}(x_{\mathrm{Mg}})=3.85\times x_{\mathrm{Mg}} \mathrm{~(kV\,cm^{-1})}.
	\end{equation}
	Given that $L_{\scriptscriptstyle W} \ll L_{\scriptscriptstyle B}$, we obtain $ F_{\scriptscriptstyle W} \simeq F_{\mathrm{max}}=632 \mathrm{~kV\,cm^{-1}}$.	
	
	The evolution of the $e_1hh_1$ exciton transition energy as a function of $L_{\scriptscriptstyle W}$ was computed. We found a value of 3.455~eV for $L_{\scriptscriptstyle W}=9~\mathrm{ML}$ and a value of 3.439~eV for $L_{\scriptscriptstyle W}=10~\mathrm{ML}$, in good agreement with the transition energies observed in reflectance and PL spectra, respectively.
	
	The PL spectrum of the QW exciton does not display any phonon replicas but features a lower-energy companion at 3.42~eV. Intensity-dependent studies (not shown here) suggest that this low-energy satellite arises from the recombination of a biexciton.
	
	\subsection{Exciton lifetime}\label{Exciton lifetime}
	
\begin{figure}[h!]
		\includegraphics[width=9cm]{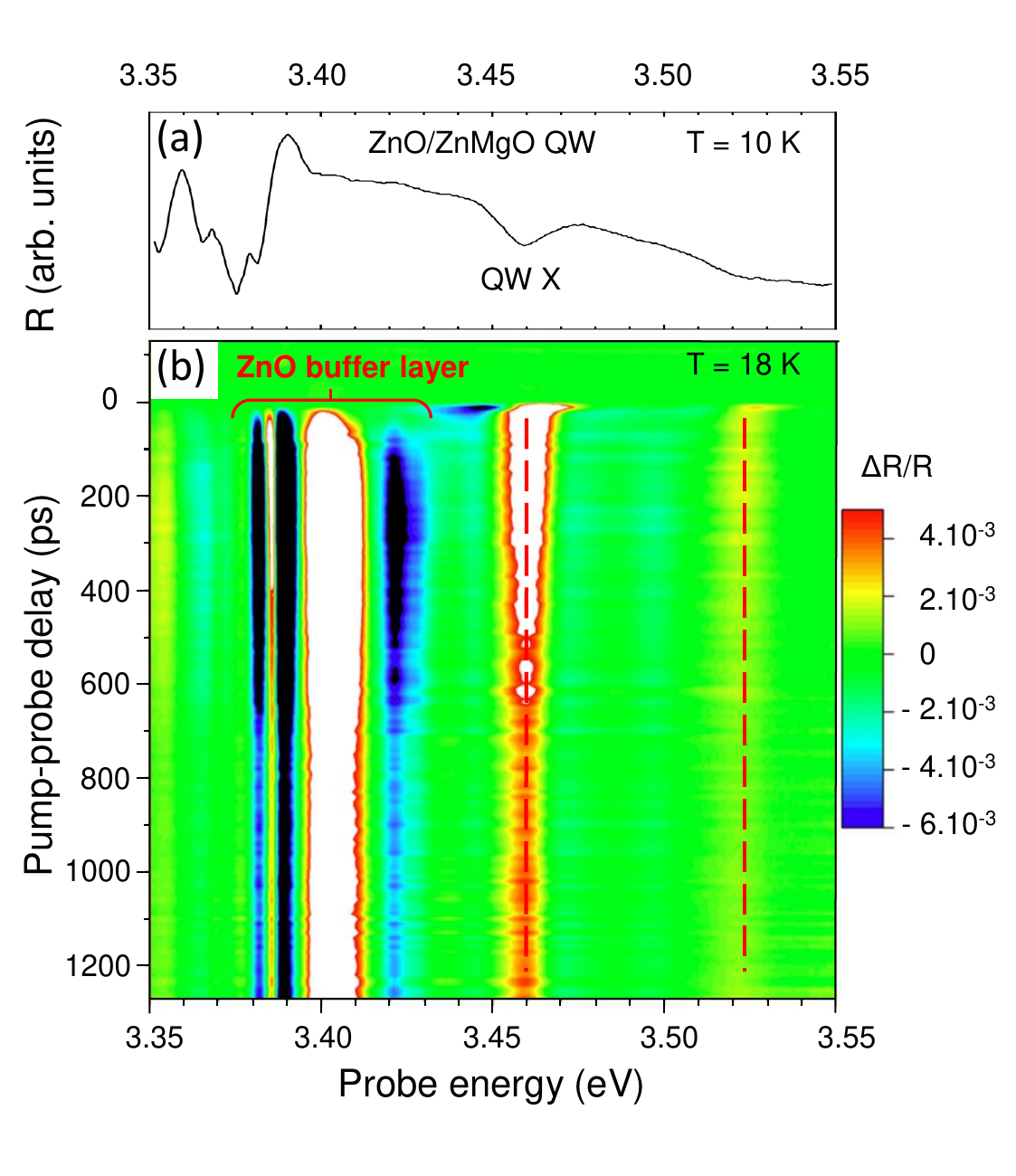}
		\caption{Linear reflectance spectrum (a) and differential reflectance as a function of both the probe energy and the pump-probe delay (b). See the detailed transition assignments in the main text.}
		\label{fig:Map 2D}
	\end{figure}

	Figure~\ref{fig:Map 2D} presents the intensity of the differential reflectance $\Delta R/R$ as a function of the probe energy for pump-probe delays up to 1~ns. In this experiment, carriers are photogenerated at 4.66~eV, significantly above the ZnMgO bandgap energy (3.7~eV). Thanks to the broad spectral range of the supercontinuum probe, we simultaneously monitor all transitions associated with the narrow QW.
	
	The linear reflectance spectrum is also plotted for comparison.
	
	Transitions occurring at photon energies below 3.41~eV correspond to excitons in the ZnO buffer layer and are not considered here. Instead, we focus on the dynamics observed in the 3.450–3.522~eV energy range, corresponding to transitions associated with the QW. The signals around 3.455~eV match the structure observed in the linear reflectance spectrum in Fig.~\ref{fig:PL}, attributed to the fundamental exciton confined in the QW. This signal exhibits a positive component at higher energy and a negative component at lower energy. At very short delays, it displays a broader spectral shape, extending towards both higher and lower energies. It then gradually narrows over the first 10 to 20~ps before stabilizing at a fixed spectral position, which persists at later delays. This phenomenon will be discussed further below.  
	
\begin{figure}[h!]
		\includegraphics[width=8cm]{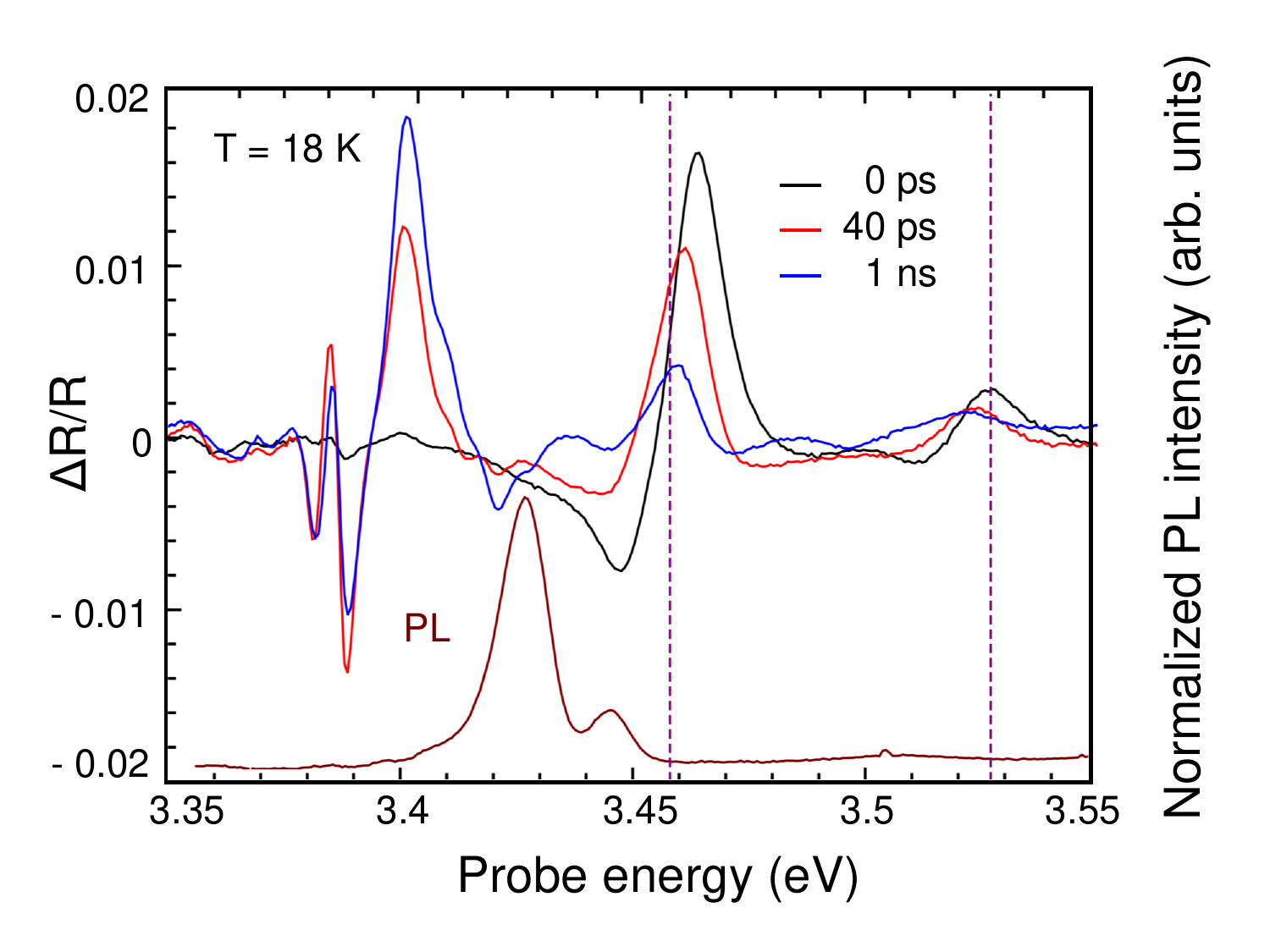}
		\caption{Differential reflectance spectra at three different pump-probe delays. The vertical dotted lines represent the transition energies in the absence of the pump.}
		\label{fig:Red shift}
	\end{figure}

	This effect is highlighted in Fig.~\ref{fig:Red shift}, which shows the $\Delta R/R$ spectra for three different pump-probe delays. The two vertical dashed lines indicate the transition positions observed in the linear reflectance spectrum, with the PL spectrum plotted for comparison.
	
	The positive component of the signal observed at the exciton energy primarily originates from bleaching due to phase-space filling. This is attributed to the presence of free excitons. The same figure also reveals that no specific PL signal is detected at the energy of the free QW exciton. This suggests that it does not decay through radiative recombination but instead thermalizes, feeding the population of localized excitons and their associated complexes. Given that no PL emission is observed at the energy of the free exciton and that its lifetime is significantly longer than the expected thermalization time, we deduce that, at least at low temperatures, the radiative lifetime of the localized exciton governs the dynamics of the free exciton.  
	
	\begin{figure}[h!]
		\includegraphics[width=11cm]{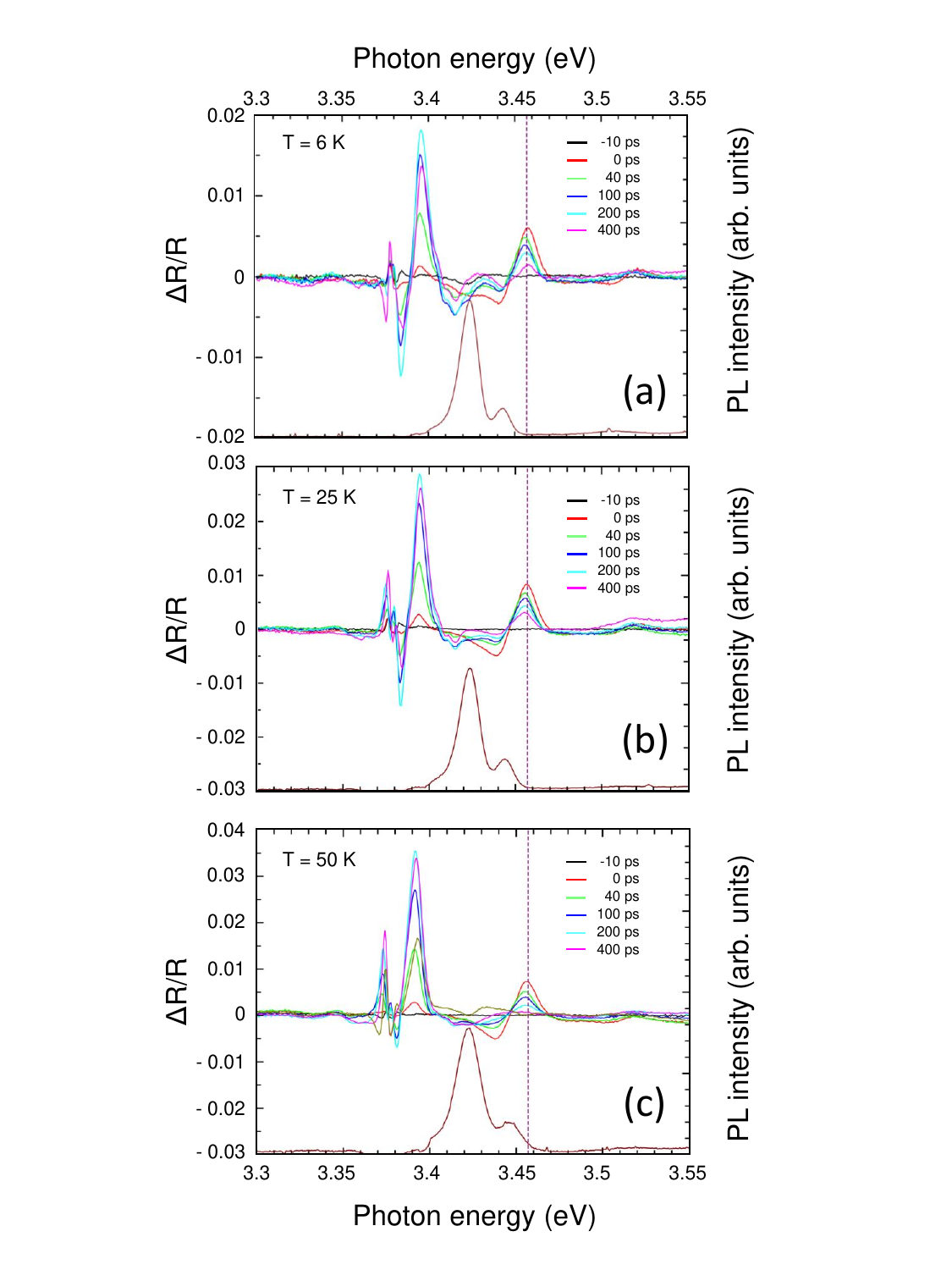}
		\caption{Comparison between $\Delta R/R$ spectra at different delays and the PL spectrum at T = 6 K (a), T = 25 K (b), and T = 50 K (c). The dotted line marks the optical transition from the linear reflectance.}
		\label{fig:3templow}
	\end{figure}
	
\begin{figure}[h!]
		\includegraphics[width= 11cm]{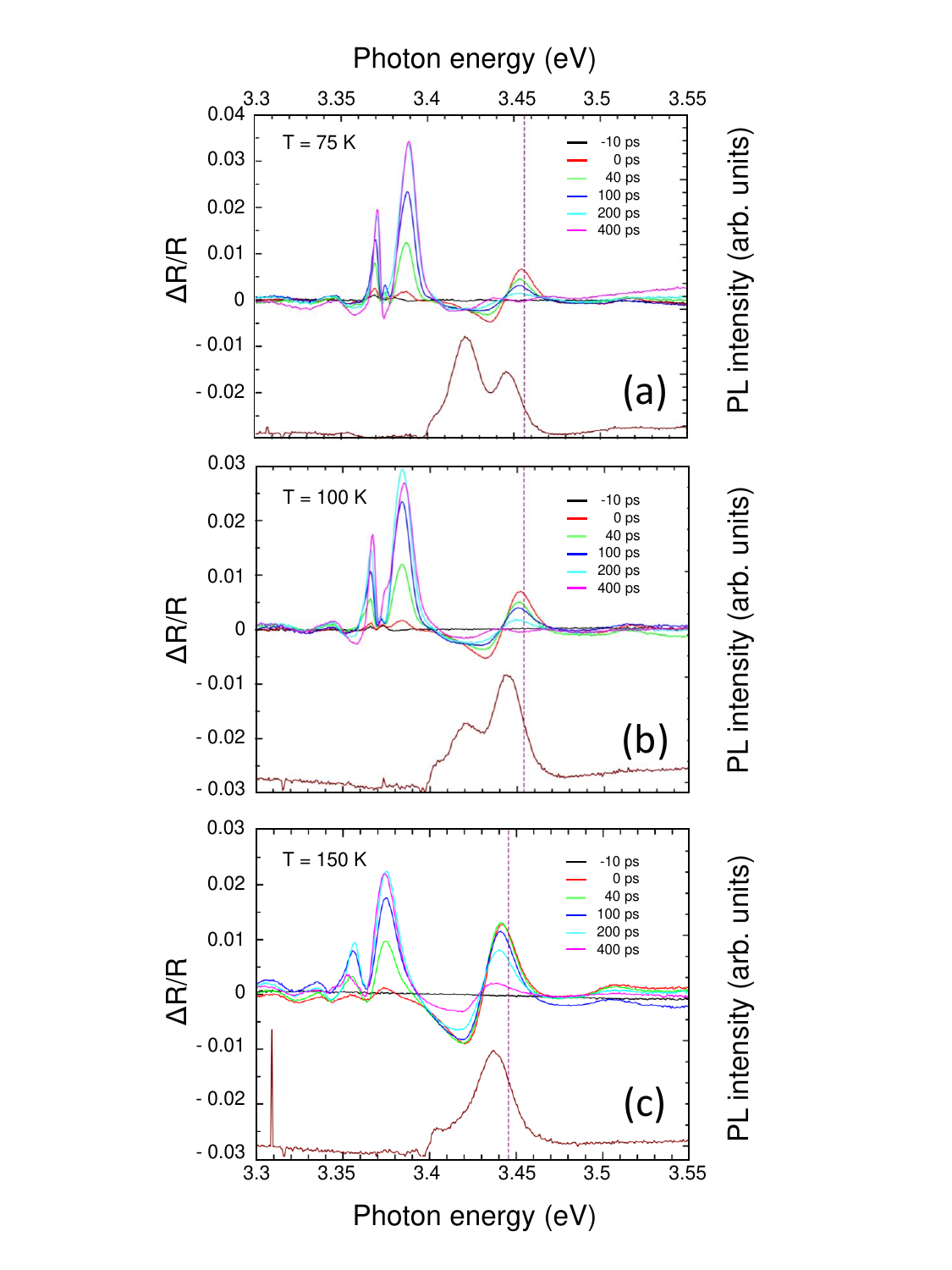}
		\caption{Comparison between $\Delta R/R$ spectra at different delays and the PL spectrum at T = 75 K (a), T = 100 K (b), and T = 150 K (c).}
		\label{fig:3temphigh}
	\end{figure}

	To further investigate this point, we varied the sample temperature. The results, compared with PL spectra measured at similar temperatures, are presented in Fig.~\ref{fig:3templow} and Fig.~\ref{fig:3temphigh}.
	
	Figs.~\ref{fig:3templow} and ~\ref{fig:3temphigh} show $\Delta R/R$ spectra at different delays compared with the PL spectrum for six temperatures (6, 25, 50, 75, 100, and 150 K). When the differential signal is compared to the PL spectrum in the 6–50~K temperature range, the localized exciton is observed in the PL spectrum, while the free exciton is identified in the differential reflectance signal. As the temperature increases, the excitons responsible for PL gradually delocalize. The PL maximum shifts towards higher energies and approaches the energy of the structure observed in $\Delta R/R$, as shown in Fig.~\ref{fig:3templow}. 
	
	As the temperature further increases from 75~K to 150~K, the energy difference between the two spectra continues to diminish, eventually leading to the alignment of the $\Delta R/R$ signal and the PL exciton line at 3.44~eV. This value is consistent with the free exciton energy previously measured and calculated, taking into account the variation of the ZnO bandgap with temperature. At this temperature, only free excitons are present, as confirmed by the evolution of the bleaching signal dynamics with temperature.
	
	\begin{figure}[h!]
		\includegraphics[width= 8cm]{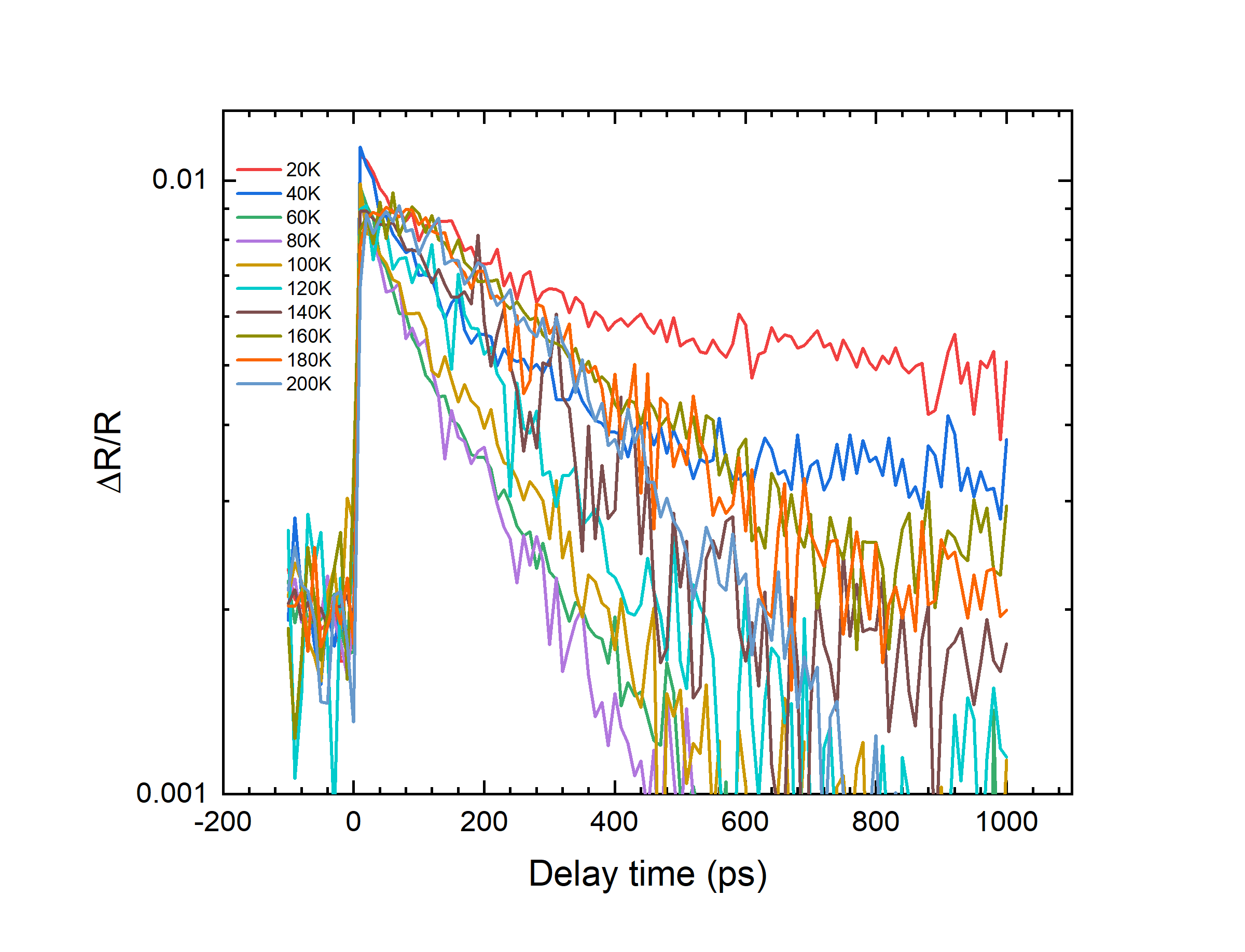}
		\caption{$\Delta R/R$ signal dynamics at the exciton photon energy for temperatures between 20 and 200 K.}
		\label{fig:templifetime}
	\end{figure}
	
\begin{figure}[h!]
		\includegraphics[width= 8cm]{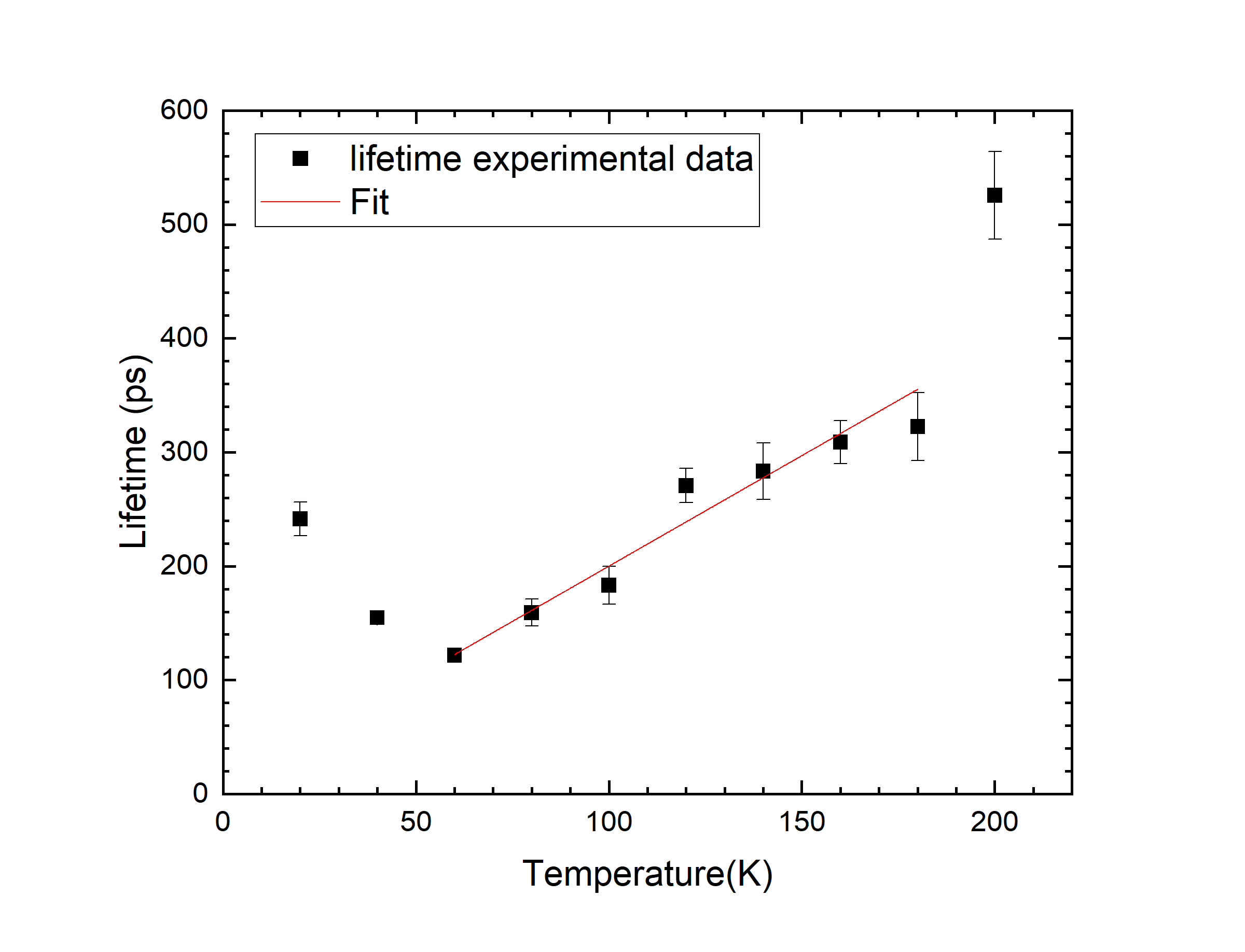}
		\caption{Evolution of the free exciton relaxation time as a function of temperature (black squares). The red line is a linear fit.}
		\label{fig:lifetime}
	\end{figure}

	Decay curves of the free exciton for temperatures ranging from 20~K to 200~K are shown in Fig.~\ref{fig:templifetime}. The lifetimes obtained from exponential fits are plotted in Fig.~\ref{fig:lifetime} for temperatures between 6 and 150~K.
	
	The decrease in lifetime from 250 to 100~ps observed between 20 and 60~K suggests that, at low temperatures, the exciton lifetime is increased due to localization within the QW plane. This phase corresponds to the progressive delocalization of the exciton population, leading to a decrease in lifetime. This process continues until 60~K, where the entire exciton population becomes delocalized. Beyond this temperature, a linear increase in lifetime is observed, as expected for free QW excitons, with the slope inversely proportional to the exciton oscillator strength~\cite{Ivanov1999}.
	
	We have fitted this lifetime evolution using a linear function and obtained:
	\[
	T_{\mathrm{X}}\left[\mathrm{ps}\right] = 1.9\,T\left[\mathrm{K}\right] + 6.0
	\]
	From this, we extrapolate a free exciton recombination lifetime of $\mathrm{6~ps \pm 1.2~ps}$ at 0~K. This result supports our earlier hypothesis that, at low temperatures, the free exciton dynamics is governed by the recombination time of localized excitons.
	
	Let us now focus on the structure that appears faintly at approximately 3.522~eV in the linear reflectance spectrum. It becomes more intense and sharper in the $\Delta R/R$ spectrum and is observed at all temperatures (Figs.~\ref{fig:3templow} and~\ref{fig:3temphigh}). This feature exhibits a spectral shape similar to that of the free exciton and follows the same temporal dynamics.
	
	\begin{figure}[h!]
		\includegraphics[width= 8cm]{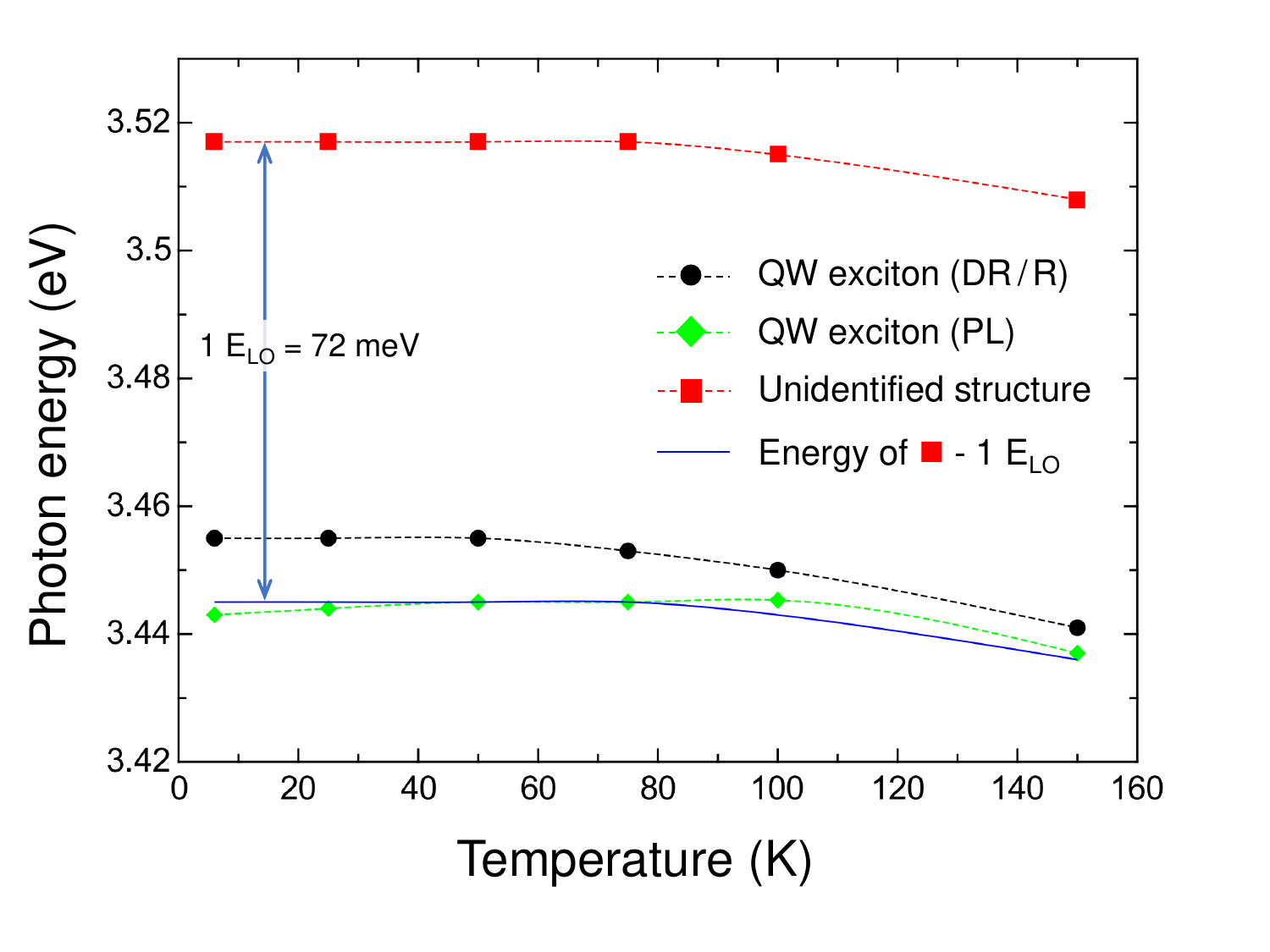}
		\caption{Temperature dependence of the QW exciton energies measured from the $\Delta R/R$ spectrum (black dots) and the PL spectrum (green diamond markers), along with the energy of the unidentified structure observed in the $\Delta R/R$ spectrum (red squares). The solid blue line represents the energy of the unidentified transition reduced by one LO-phonon energy (72~meV).}
		\label{fig:Phonon replica}
	\end{figure}
	
	Figure~\ref{fig:Phonon replica} shows the temperature dependence of the quantum well exciton energies measured in both $\Delta R/R$ and PL, compared to the evolution of the energy of the unidentified structure. The solid blue curve represents the energy of this structure reduced by the longitudinal optical (LO) phonon energy in ZnO ($E_{\mathrm{LO}} = 72$~meV). It closely follows the evolution of the PL exciton energy, suggesting that the observed structure originates from phonon-assisted exciton creation as observed by Beaur \textit{et al}.\cite{Beaur2013} in similar samples.
	
	However, the fact that this structure lies exactly 72~meV above the PL peak, and not above the excitonic feature in the $\Delta R/R$ spectrum, indicates that it involves the direct creation of a localized exciton, rather than a free exciton. Furthermore, since its temporal dynamics match those of the free exciton bleaching, this implies that, as in the case of the free exciton, the decay is governed by the lifetime of the final state — namely, the thermalized exciton. This is not unexpected, as the interaction time is on the order of 100~fs, making the relaxation toward the thermalized state effectively instantaneous on the timescale of our experiment.
	
	\subsection{Biexciton}
	At all temperatures and pump-probe delays, the positive exciton bleaching signal is accompanied by a negative signal. A negative signal characterizes a pump-induced transition from the excitonic state: the pump pulse populates a state from which a probe photon is absorbed, leading to a new state. In this case, it corresponds to the creation of a biexciton~\cite{kleinman1983} from the exciton population generated by the pump.
	
	\begin{figure}[h!]
		\includegraphics[width= 7cm]{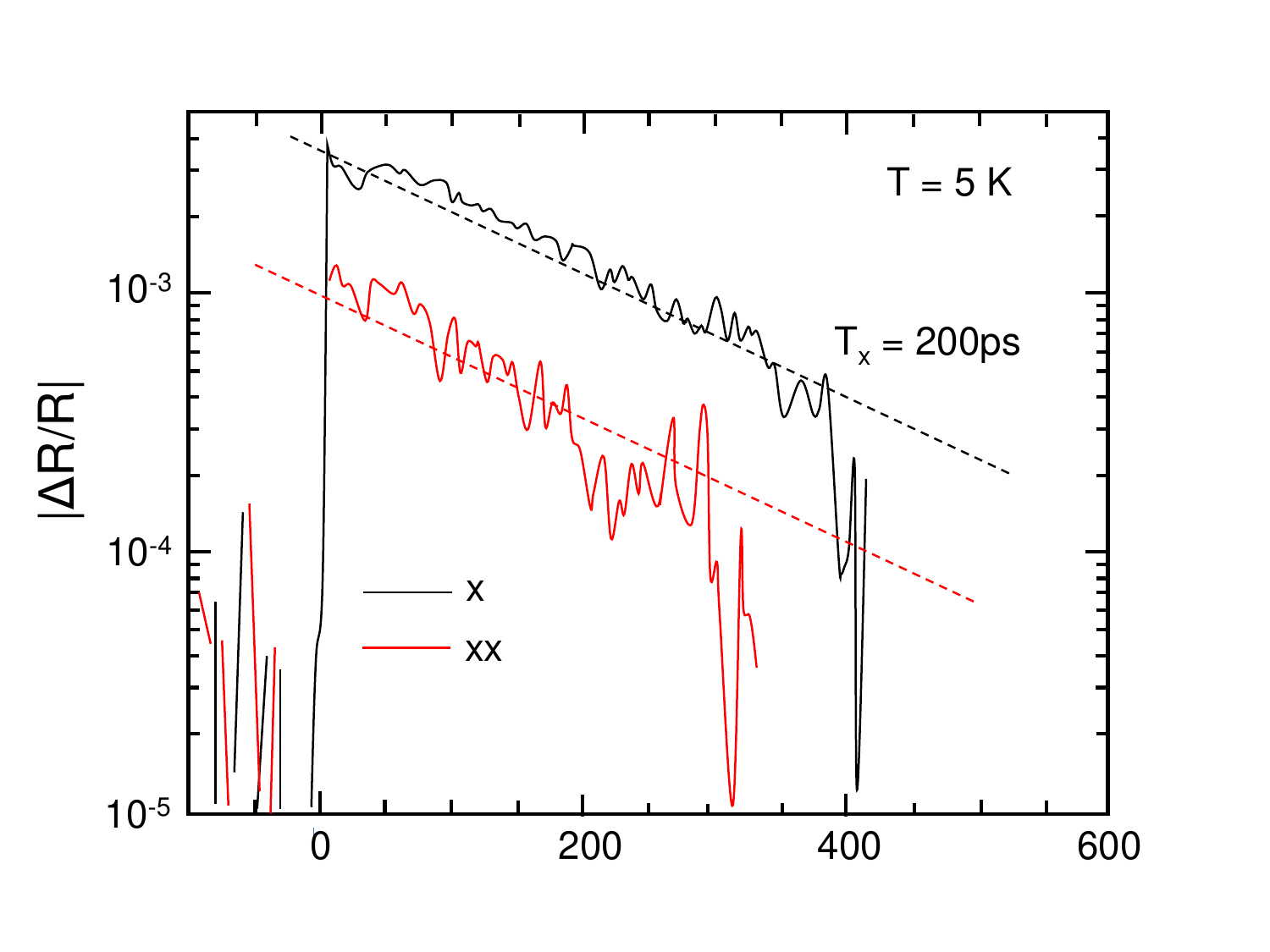}
		\caption{Times evolution of both the positive signal~(x) and the negative one (xx), at the excitonic resonance. The positive signal is associated with the exciton bleaching while the negative one is interpreted as an induced absorption towards the biexciton.}
		\label{fig:Decays x and xx}
	\end{figure}
	
	We will now confirm this interpretation by analyzing both the dynamics and the spectral shape of the negative signal.
	
	From a dynamical perspective, if the transition associated with this signal indeed corresponds to the absorption of a probe photon from the population of excitons created by the pump, then it must be proportional to the exciton population and exhibit the same temporal evolution. This is precisely what is observed in Fig.~\ref{fig:Decays x and xx}, where the temporal profiles of the bleaching and negative signals are plotted. Note that the absolute value of the negative signal is shown in order to allow a direct comparison of the two dynamics. Both curves follow an exponential decay with a characteristic time of 200 ps.
	
	In addition to this temporal analysis, further insight can be gained by examining the spectral shape of the negative signal. The ratio of effective masses between the biexciton and the exciton, along with the momentum conservation rule for photons, implies that the spectral shape of the induced signal at lower energies must reflect the thermal occupation of excitonic states~\cite{Gourley1982}. Specifically, the spectral profile of the emission band results from the convolution of a Boltzmann distribution with the joint density of states of the exciton and the biexciton.  
	
\begin{figure}[h!]
		\includegraphics[width= 8cm]{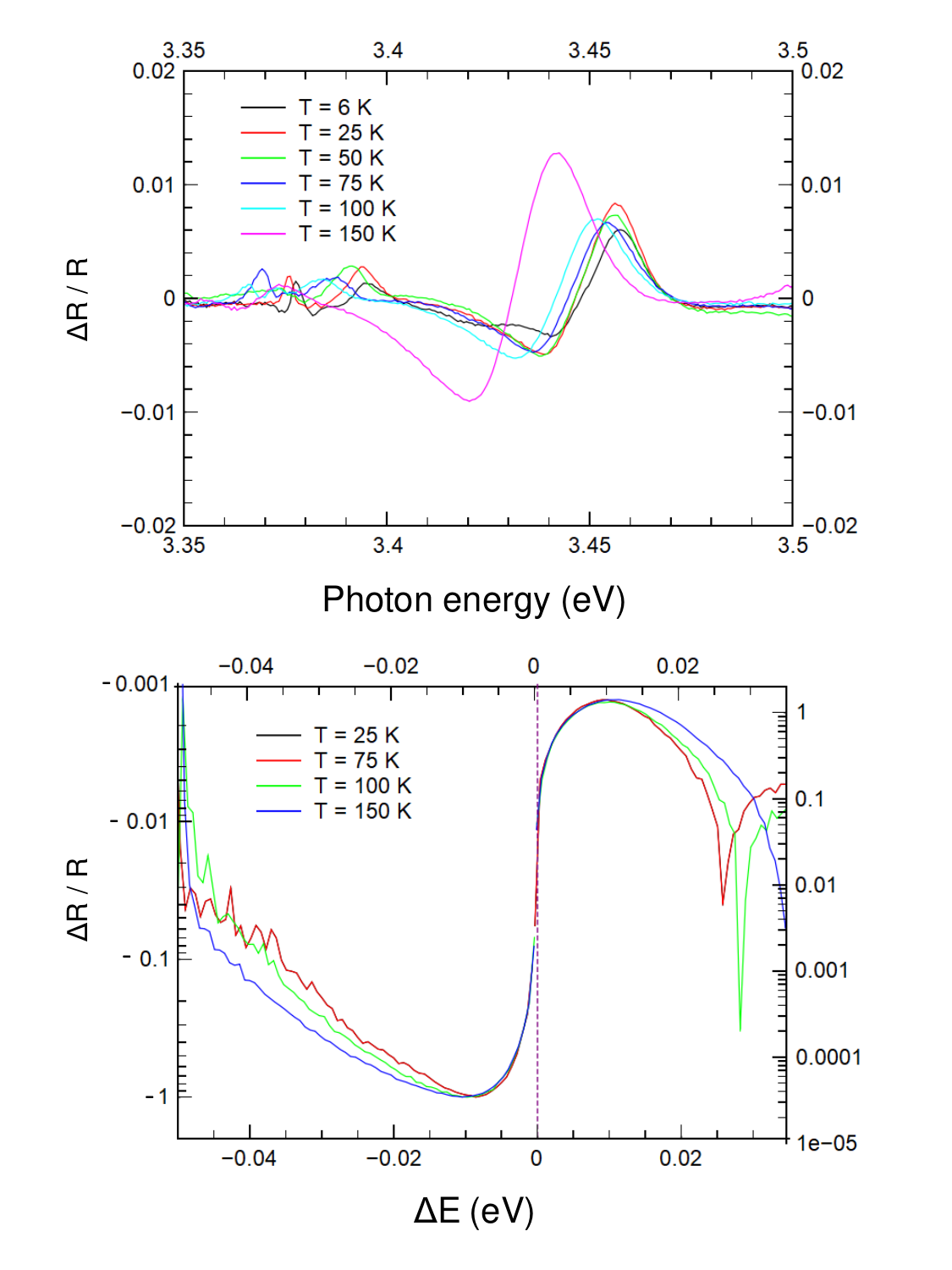}
		\caption{(a) Evolution with temperature of the $\Delta R/R$ signal at the excitonic resonance. (b) Same spectra, normalized and superimposed for the larger temperature plotted as a function of the difference with the energy of the structure inflection point. (Note the different scales used on the two sides of graph on the right panel for negative and positive signals.)}
		\label{fig:XXiden}
	\end{figure}

	To verify this hypothesis, we analyzed the shape of the $\Delta R/R$ spectrum at the exciton energy as a function of temperature. Figure~\ref{fig:XXiden}(a) presents the differential reflectance spectra recorded 10~ps after the arrival of the pump pulse for the six studied temperatures. 
	
	To facilitate comparison, the spectra were normalized and superimposed such that the inflection point of the structure aligns at the same energy for all six spectra.  
	We then plotted these spectra on a semilogarithmic scale as a function of the energy deviation relative to the axis of antisymmetry of the spectrum, as shown in Fig.~\ref{fig:XXiden}(b). For clarity, only three temperatures are displayed in this representation, as variations were found to be negligible at lower temperatures.

	We find that as the temperature increases, the signal broadens both towards higher energies for the exciton (positive signal) and towards lower energies for the biexciton (negative signal). The broadening of the exciton signal at higher energies is attributed to the progressive filling of excitonic states. This state filling also indirectly causes the broadening of the biexciton signal towards lower energies, as the excitonic level serves as the initial state for this transition.  
	
	The biexciton binding energy can be estimated by taking the energy difference between the exciton and biexciton peaks. We obtain a value ranging from 18 to 22~meV, depending on temperature. This variation results from the redshift of the biexciton energy~\cite{Levy2001, Davis2009}, while the exciton energy shifts towards higher values as the temperature increases. 
	
	Let us now revisit a point previously deferred and introduced in Section~\ref{Exciton lifetime}: in the early sub-picosecond regime, the exciton and biexciton signals both exhibit pronounced spectral broadening. We first focus on the positive exciton-related signal, whose broadening is attributed to a hot exciton distribution, strong carrier–carrier scattering, and transient screening effects immediately following photoexcitation. As the system evolves during the first few picoseconds, these interactions weaken as the exciton population cools down and relaxes toward the band edge, resulting in a progressive narrowing of the exciton signal.
	
	Turning to the biexciton-related negative signal, we emphasize that it originates from a photoinduced absorption process involving the excitonic state as the initial state. Consequently, it does not reflect the intrinsic lifetime of biexcitons, but instead follows the exciton population dynamics. Its spectral width is primarily governed by the energy distribution of the occupied excitonic states and may also be affected by variations in biexciton binding energies. This interpretation is reinforced by the temperature-dependent analysis presented earlier, which shows that exciton state filling at elevated temperatures causes a broadening of the exciton signal toward higher energies. As a direct consequence of this, the biexciton-related signal simultaneously broadens toward lower energies, consistent with the excitonic origin of the initial state involved in the transition.

	%\PGremove{\section{Discussion}}
\subsection{Biexciton binding energy}
\begin{figure}[h!]
		\includegraphics[width=8cm]{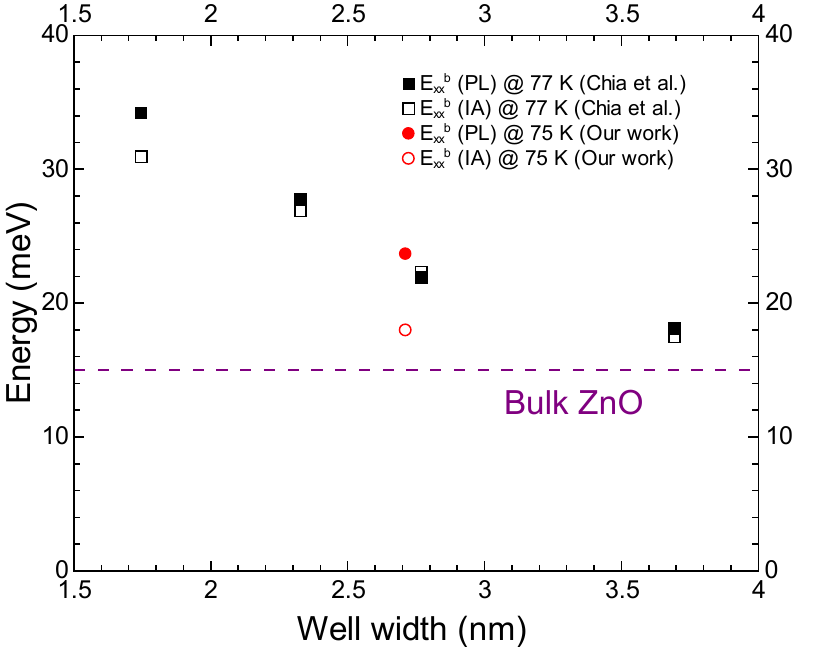}
		\caption{Binding energy of localized and free biexcitons determined is this work by Pl and induced absorption (IA), respectively. Data of Ref.~\cite{Chia2001} are also plotted for comparison.}
		\label{fig:summary}
	\end{figure}

	The binding energy of the biexciton in bulk ZnO is 15 meV~\cite{Klingshirn2007,Hazu2003}. In QWs, the confinement is expected to increase the binding energy. Indeed, Sun et al~\cite {Sun2001,Sun2002} determined the binding energy of localized biexcitons in $\mathrm{ZnO/Zn_{0.74}Mg_{0.26}O}$ QW with different widths using low temperature PL (T = 5 K). They measured binding energies between 19 meV and 28 meV when the well width is decreased from 3.7 nm to 1.75 nm. In addition, the same authors performed both PL and nanosecond pump-probe spectroscopy at 77 K to study series of similar samples with the same Mg fraction of 26\% but with well widths now spanning values between 1.75 and 4.23 nm~\cite{Chia2003}. They were thus able to differentiate, as we did, the binding energy of the free biexciton and the binding energy of the biexciton localized on the interface fluctuations of the QW. Their values are synthesized with ours in fig~\ref{fig:summary}.
	Biexcitons were also observed in non-polar ZnO QWs. The well thickness was 3.5 nm and the Mg barrier composition 20\%. Both power and temperature dependent PL measurements~\cite{Ali2018} enabled to estimate a binding energy of 45 meV. This value is 1.5 times higher than what was measured by Chia et al~\cite{Chia2003} in a well twice as narrow with a higher Mg concentration in the barrier. It can be assumed that in a polar well the confined quantum Stark effect is responsible for a decrease in the biexciton binding energy in the same way that it decreases the exciton binding energy. 
	With respect to the relaxation dynamics of the different transitions observed, three of them are identical. Thus, the signal associated with free excitons presents a decay time which is conditioned by the radiative duration of the localized excitons. Similarly, the signal at an energy of 3.522~eV corresponds to a phonon-assisted absorption process resulting in the creation of a thermalized exciton and exhibits exactly the same spectro-temporal dynamics as the free exciton resonance; this is due to the high efficiency of the thermalization process in both cases. It is surprising that this phenomenon is observable when no phonon replica appears clearly in the emission spectrum of the 2.6~nm thick QW even though they are particularly remarkable in the PL of both the barriers and the 7.1~nm wide QW. Finally, the induced absorption signal towards the biexciton shows the same lifetime because it is proportional to the population of localized excitons.
	
\subsection{Spin relaxation induced by exciton formation}
	We can now consider the use of polar ZnO quantum wells a potentially attractive platform for spin manipulation, a prospect that motivates current studies of this material. The spin control is here based on the presence of an internal electric field due to the quantum-confined Stark effect which induces a Rashba-type spin–orbit interaction. As theoretically predicted\cite{Harmon2011}, this interaction may partially compensate for the Dresselhaus term, leading to reduced spin decoherence through suppression of the Dyakonov–Perel mechanism. 
	
	However, our time-resolved measurements reveal certain limitations. In particular, free carriers quickly bind into free excitons, which do not recombine radiatively but instead rapidly relax into localized or thermalized states at low temperatures. This sub-picosecond thermalization shortens the effective lifetime of free excitons and can result in the loss of spin polarization through energy and momentum relaxation before recombination occurs. Although the localized exciton population exhibits well-defined lifetimes, the associated spin coherence is likely degraded during the relaxation process.
	
	Moreover, while the strong exciton binding energy stabilizes the excitonic state against thermal dissociation, it also implies a strong spatial overlap between the electron and hole wavefunctions. This enhances the exchange interaction, which can cause rapid relaxation of the exciton's total angular momentum. Such exchange-driven spin depolarization has been reported in other wide-bandgap semiconductors, such as GaN, and may similarly impact spin retention in ZnO-based quantum wells.
	
	Finally, the coexistence of multiple optical transitions — including phonon-assisted absorption and biexcitonic states — adds spectral complexity. Although it reveals rich many-body physics, it may complicate the spectral selectivity and polarization-resolved detection required for precise spin injection and readout schemes.

	\section{Conclusion}
	%Using non-conventional time-resolved pump-probe spectroscopy, we have investigated the dynamics of photocreated excitons and biexciton in a 2.6~nm thick ZnO/Zn$_{O.84}$Mg$_{0.16}$ QW. The dynamics of free exciton is governed by the radiative lifetime of the localized exciton as evidence of an efficient thermalization process. Similarly, we observed an optical phonon-assisted absorption process that directly results in the creation of a localized exciton.\PGadd{ }The existence of an absorption process of the probe induced by the presence of the pump allowed us to determine the binding energy of the biexciton, that in our experimental conditions, the biexcitons are created directly from the thermalized population. Finally, a short-lived redshift of the carrier is observed which was previously neglected. This shift could have a negative implication to spin injection efficiency hence impeding direct spin observation..
	We have performed a detailed investigation of exciton and biexciton dynamics in a ZnO/ZnMgO quantum well using time-resolved pump–probe spectroscopy. Our measurements reveal the ultrafast capture and relaxation of photogenerated carriers from the barrier into the well, followed by the rapid formation of excitons within a few hundred femtoseconds. These excitons are subsequently localized, likely due to well-width fluctuations or interface disorder, as evidenced by the absence of free exciton emission and the Stokes shift observed in photoluminescence.
	
	In addition to these localized excitons, we observe the formation of biexcitons through photoinduced absorption from the exciton population, with a binding energy ranging from 18 to 22~meV depending on temperature. A phonon-assisted absorption feature is also identified and interpreted as the creation of thermalized excitons, further highlighting the richness of the excitonic landscape in this system.
	
	The efficiency of exciton formation and the rapid thermalization of electron–hole pairs underscore the strong Coulomb interaction in these quantum wells. the biexciton formation that we observe in both PL and pump-and-probe experiments reinforce this conclusion. However, this same interaction, along with the strong spatial overlap between electrons and holes, also enhances exchange processes that can contribute to spin depolarization. Combined with the fast localization dynamics, this may limit the persistence of spin-polarized states, which are essential for optical spin injection schemes.
	
	Nonetheless, the presence of internal electric fields and Rashba-type spin–orbit coupling offers interesting possibilities for spin manipulation. Enhancing the internal piezoelectric filed could allow for better electron hole separation: that would avoid exciton formation and wave function overlap which increase the exchange interaction, detrimental for spin preservation. Tailoring the quantum well structure to control localization and exchange effects will be a key step toward enabling efficient spintronic functionalities in ZnO-based systems.
	
	\section*{Acknowledgements}	
	The authors gratefully acknowledge the support and funding of the French 'Agence Nationale de la Recherche' (ANR) under reference ANR-19-CE24-0020.
	
	%\section*{References}
	%\bibliographystyle{plain}
	\bibliography{references}

\end{document}